\documentstyle[12pt]{article}
\begin{document}
\baselineskip 18pt
\begin{titlepage}
\centerline{\Large\bf $SO(3,2)$ structure and distributions of two-component}
\centerline{\Large\bf Bose-Einstein condensate with lower excitations}
\vspace{2cm}
\centerline{ Hong-Biao Zhang $^{1,2}$, Bing-Hao Xie $^{2}$ and Mo-Lin Ge $^{2}$}
\vspace{1cm}
\centerline{\it 1. Department of Physics, Northeast Normal University,}
\centerline{\it    Changchun, Jilin, 130024, P. R. China }
\centerline{\it 2. Theoretical Physics Division, Nankai Institute of Mathematics,}
\centerline{\it    Nankai University, Tianjin 300071, P. R. China}
\vspace{2cm}

\centerline{\bf Abstract}
 The wave function describing two-component Bose-Einstein
 condensate with weakly excitations has been found, by using the
 SO(3,2) algebraic mean-field approximation. We show that the two-component
 modified BEC (see eq.(\ref{ga})) possesses uniquely super-Poissonian
 distribution in a fixed magnetic field along $z$-direction. The distribution
 will be uncertain, if $\bf {B}=0$.

\vspace{1cm}


Electronic address: hbzhang@eyou.com

\vspace{3cm}

\pagebreak

\section{Introduction}

 Bose-Einstein condensation (BEC) was first introduced as a phenomenon
 in non-interacting Bose system and was generalized to
 interacting Bose systems in 1956 by Penrose and Onsager\cite{pen}.
 More recently, following the experiments with BEC of atomic
 gases\cite{and1}-\cite{dav}, simultaneous
 condensation of two different atomic species or two different
 hyperfine spin states of the same atoms such as $^{87}R_{b}$
 was achieved in the same trap\cite{mya}-\cite{hall2} that
 has further stimulated a great revival of
 interest in the theoretical study of this phenomenon.
 The multiplicit BEC theory was first set up by Ho\cite{ho}.
 One fascinating aspect of BEC is the nature of coherence for
 a macroscopic quantum system, and in recent experiments some
 of the coherence properties of BEC have also been discussed and
 explicitly addressed\cite{and2}-\cite{burt}. Among number of papers
 the algebraic average method (AAM) was used to discussed
 one-component BEC by Solomon\cite{solo} et al. However, many
 properties of two-component (or binary) BEC may still be
 desirable to be understood from the coherence.
 In comparison with one-component macroscopic quantum Bose system, the physics
 of binary BEC is richer than of the usual one-component systems.
 if the magnitude of wavefunction may not be constant
 the advantage of AAM appears. In this paper, we extend the idea
 in Ref.[11] to two-component case and describe the Hamiltonian
 and energy eigenstate within the SO(3,2) mean-field picture of BEC.
 Based on this theory, a generalized version of the BEC
 weakly excited states is constructed. The second-ordered
 correlation functions is also calculated.

 \section{Model and solution}

 The standard description of two-component Bose-Einstein condensation
 is by means of two-component bosonic atomic fields
 $\Psi_{\alpha}({\bf x})$ $(\alpha=1,2)$ and thus the density
 of the particle number and the spin can be written as
 $n({\bf x})=\sum_{\alpha}\Psi^{\dag}_{\alpha}
 ({\bf x})\Psi_{\alpha}({\bf x})$, and
 ${\bf S}({\bf x})=\sum_{\alpha\beta} \Psi^{\dag}_{\alpha}({\bf x})
 {\bf S}_{\alpha\beta}\Psi_{\beta}({\bf x})$, respectively.
 In the presence of a constant magnetic field ${\bf B}$
 along the $z$-direction the Hamiltonian take the form\cite{ohmi}:
\begin{eqnarray}
\label{1.1}
   H &=& \int d^3 x \{\sum_{\alpha}
     \Psi^{\dag}_{\alpha}({\bf x})
   (-\frac{{\nabla}^2}{2M}+U({\bf x}))\Psi_{\alpha}({\bf x})
   +\frac{1}{2}g_{n} n^{2}({\bf x}) \nonumber \\
  & & +\frac{1}{2}g_{s}{\bf S}({\bf x})\cdot{\bf S}({\bf x})
   -g_{\mu}{\bf B}\cdot{\bf S}({\bf x})\}
\end{eqnarray}
where $g_{n}$, $g_{s}$ are coupling constants and $g_{\mu}$ is the gyromagnetic
ratio. Expanding $n({\bf x})$ and ${\bf S}({\bf x})$ in
terms of the field operators, eq.(\ref{1.1}) can be expressed as
\begin{eqnarray}
\label{1.2}
H &=& \sum_{\alpha} \int d^3 x \Psi^{\dag}_{\alpha}({\bf x})
   (-\frac{{\nabla}^2}{2M}+U({\bf x})+g_{1})\Psi_{\alpha}({\bf x})
   \nonumber \\
& &-\frac{1}{2}g_{\mu}B\int d^3 x (\Psi^{\dag}_{1}({\bf x}) \Psi_{1}({\bf x})
- \Psi^{\dag}_{2}({\bf x}) \Psi_{2}({\bf x})\} \nonumber \\
& &+\frac{1}{2}g_{2} \sum_{\alpha\beta} \int d^3 x
   \Psi^{\dag}_{\alpha}({\bf x}) \Psi^{\dag}_{\beta}({\bf x})
  \Psi_{\beta}({\bf x}) \Psi_{\alpha}({\bf x})
\end{eqnarray}
 where $g_{1}=\frac{1}{2}{g_n}+\frac{3}{8}{g_s}$ and
 $g_{2}=g_{n}+\frac{1}{4}{g_s}$.
In two dimensions in the cylindrical polar coordinates through the
Fourier transformation
$$\Psi_{1}({\bf x})={\sum}_{m}a_m \Phi_{m}({\bf x}),\;\;\;
  \Psi_{2}({\bf x})={\sum}_{m}b_m \Phi_{m}({\bf x})$$
 with
 $$\Phi_{m}({\bf x})=\frac{1}{({\pi}^{\frac{3}{2}}m!a^3_{osc})^{\frac{1}{2}}}
 (\frac{\rho}{a_{osc}})^{|m|}e^{im\phi}e^{\frac{-({\rho}^2+z^2)}{2a^{2}_{osc}}}
$$
being the harmonic-oscillator wave functions.
$a_{osc}=(\frac{\bar{h}}{M\omega})^{\frac{1}{2}}$
is the harmonic oscillator length, further assume that the
bosons are in their ground state with respect to
the $z$-axis, it is recast eq.(\ref{1.2}) to the
second quantized form\cite{steel}
\begin{eqnarray}
\label{e1}
  H &=& \sum_{k} \{\epsilon_{k} (n_{k}^a +  n_{k}^b)
  +\frac{1}{2}g_{\mu}B(n_{k}^{b}-n_{k}^{a})\} \nonumber \\
 & & +\frac{1}{2}g_{2}\sum_{{k},{l},{m},{n}} \langle
 {k,l}\mid{m,n}\rangle
 (a^+_{k}a^+_{l}a_{m}a_{n}
 + b^+_{k}b^+_{l}b_{m}b_{n}
 +2a^+_{k}b^+_{l}a_{m}b_{n})
\end{eqnarray}
where
\begin{eqnarray*}
 \langle{k,l}\mid{m,n}\rangle
&=& \int \Phi_{k}^{*}({\bf x}) \Phi_{l}^{*}({\bf x})
 \Phi_{m}({\bf x}) \Phi_{n}({\bf x}) d^{3}{x} \\
&=& \delta_{k+l,m+n}\frac{(k+l)!}{2^{(k+l)}\sqrt{k!l!m!n!}}V_{0} \\
V_{0} &=& \int\mid\Phi_{0}({\bf x})\mid^{4}d^{3}x
\end{eqnarray*}
The number operators $n_k^a \equiv a^+_ka_k$ and
$n_k^b \equiv b^+_kb_k$, the raising operators
$a^+_k$ ($b^+_k$), and the lowering operators
$a_p$ ($b_p$) obey the Weyl-Heisenberg algebraic
commutators:
\begin{eqnarray}
\label{e2}
& & [a_p,a^+_k]=\delta_{pk} ,\;\;\;
 [n_k^a,a^+_p]=\delta_{pk}a^+_k ,\;\;\;
 [n_k^a,a_p]=-\delta_{pk}a_k , \nonumber \\
& & [a_p,b^+_k]=[a_p,b_k]=[b_p,a^+_k]=[b^+_p,a^+_k]=0 \nonumber \\
& & [b_p,b^+_k]=\delta_{pk} ,\;\;\;
[n_k^b,b^+_p]=\delta_{pk}b^+_k ,\;\;\;
[n_k^a,a^+_p]=-\delta_{pk}b_k .
\end{eqnarray}
The Bogoliubov prescription is that at zero temperature the state with $k=0$
is macroscopically occupied and this observation allows one to treat $a^+_0$
($b^+_0$) and $a_0$ ($b_0$) as $c$ numbers ($[a_0,a^+_0]\simeq0$) since the
corresponding number operator $n^a_0$ ($n^b_0$) respectively, counting the
bosons constituting the condensate, turns out to be macroscopically large.
However, this neglect of the operator $a^+_0$ ($b^+_0$) and $a_0$ ($b_0$)
is not an appropriate approximation if we wish to describe phenomena in
the condensate ground states. So here we no longer adopt such an approximation
and we retain the operator status of $a^+_0$ ($b^+_0$) and $a_0$ ($b_0$)
in order to give a more consistent description of the state of the condensated
system.

Making explicit the terms depending on $a^+_0$ ($b^+_0$) and $a_0$ ($b_0$)
in eq.(\ref{e1}) and neglecting those terms that contain three or four boson
operators $a^+_k$ ($b^+_k$) and $a_l$ ($b_l$) ($k,l\not=0$) the $H$ reduces
to the form
\begin{eqnarray}
\label{e3}
H &=& \epsilon_{0} (n_{0}^a +  n_{0}^b)
   +\sum_{k\not=0} \epsilon_{k} (n_{k}^a  + n_{k}^b)
+\frac{1}{2} g_{2}V_{0}({a_{0}^{+}}^{2}a^2_{0}
+ {b_{0}^{+}}^{2}b^2_{0} \nonumber \\
& & + 2a^+_{0}b^+_{0}b_{0}a_{0})
 +\frac{1}{2}g_{2}V_{0}\sum_{k\not=0}\frac{1}{2^k} (4n_{0}^a n_{k}^a
 + 4n_{0}^b n_{k}^b \nonumber \\
& & + 2n_{0}^a n_{k}^b
 +2n_{0}^b n_{k}^a
 + 2a^+_{0}b_{0}b^+_{k}a_{k}
 + 2b^+_{0}a_{0}a^+_{k}b_{k}) \nonumber \\
& & + \frac{1}{2}g_{2}V_{0}\sum_{k\not=0} \frac{1}{\sqrt{k!(-k)!}}
  ({a_{0}^+}^2 a_{-k}a_{k}
 + a^2_{0}a^+_{k}a^+_{-k} \nonumber \\
& & + {b_{0}^+}^2b_{-k}b_{k}
 + b^2_{0}b^+_{k}b^+_{-k}
 + 2a^+_{0}b^+_{0}a_{k}b_{-k}
 + 2a_{0}b_{0}a^+_{k}b^+_{-k}) \nonumber \\
& & +\frac{1}{2}g_{\mu}B(n^{b}_{0}-n^{a}_{0})
+\frac{1}{2}g_{\mu}B\sum_{k\not=0} (n_{k}^b - n_{k}^a)
\end{eqnarray}

We now define the generators of the algebra $SO(3,2)$
in the following forms
\begin{eqnarray}
\label{e4}
& & E_{+}^{(0)} =\frac{1}{\sqrt 2} a^+_{0}b^+_{0}
\;,\;\;\;\;\;\;
E_{-}^{(0)} =\frac{1}{\sqrt 2} a_{0}b_{0}
\;,\;\;\;\;\;\;
 E_{3}^{(0)} =\frac{1}{2}(n^a_{0}+n^b_{0}+1)
\nonumber \\
& &  F_{+}^{(0)} =\frac{1}{\sqrt 2} a_{0}b^+_{0}
\;,\;\;\;\;\;\;
 F_{-}^{(0)} =\frac{1}{\sqrt 2} a^+_{0}b_{0}
\;,\;\;\;\;\;\;
 F_{3}^{(0)} =\frac{1}{2} (n^b_{0}-n^a_{0})
\nonumber \\
& &  U_{+}^{(0)} =\frac{1}{2} {a^+_{0}}^2
\;,\;\;\;
 U_{-}^{(0)} =\frac{1}{2} {a_{0}}^2
\;,\;\;\;
 V_{+}^{(0)} =\frac{1}{2} {b^+_{0}}^2
\;,\;\;\;
 V_{-}^{(0)} =\frac{1}{2} {b_{0}}^2
\end{eqnarray}
and
\begin{eqnarray}
\label{e5}
 & & E_{+}^{(k)} =\frac{1}{\sqrt 2}(a^+_{k}b^+_{-k}
 +a^+_{-k}b^+_{k}) \;,\;\;\;\;\;
 F_{+}^{(k)} =\frac{1}{\sqrt 2}(a_{k}b^+_{k}
 +a_{-k}b^+_{-k})
 \nonumber \\
 & & E_{-}^{(k)} =\frac{1}{\sqrt 2}(a_{k}b_{-k}
 +a_{-k}b_{-k}) \;,\;\;\;\;\;
  F_{-}^{(k)} =\frac{1}{\sqrt 2} (a^+_{k}b_{k}
 +a^+_{-k}b_{-k})
\nonumber \\
& & E_{3}^{(k)} =\frac{1}{2}(n^a_{k}+n^a_{-k}
 +n^b_{k}+n^b_{-k}+2)
 \nonumber \\
 & & F_{3}^{(k)} =\frac{1}{2} (n^b_{k}+n^b_{-k}
 -n^a_{k}-n^a_{-k})
 \nonumber \\
 & & U_{+}^{(k)} = a^+_{-k}a^+_{k}
 \;,\;\;\;\;\;\;\;\;
 V_{+}^{(k)} = b^+_{-k}b^+_{k}
 \nonumber \\
 & & U_{-}^{(k)} = a_{k}a_{-k}
 \;,\;\;\;\;\;\;\;\;
 V_{-}^{(k)} = b_{k}b_{-k}
\end{eqnarray}
which generate a $SO(3,2)$ algebra with the generators
satisfying the following relations ($q=0,k,-k$):
\begin{eqnarray}
\label{e6}
& &  [E_{\pm}^{(q)}, V_{\mp}^{(q)}]= \mp F_{\mp}^{(q)}
\;\;\;\;\;\;\;\;\;\;
  [F_{\pm}^{(q)}, V_{\mp}^{(q)}]= \mp E_{\mp}^{(q)}
\nonumber \\
& &  [E_{\pm}^{(q)}, U_{\mp}^{(q)}]= \mp F_{\pm}^{(q)}
\;\;\;\;\;\;\;\;\;\;
  [F_{\pm}^{(q)}, U_{\pm}^{(q)}]= \pm E_{\pm}^{(q)}
\nonumber \\
& &  [E_{\pm}^{(q)}, F_{\pm}^{(q)}]= \mp V_{\pm}^{(q)}
\;\;\;\;\;\;\;\;\;\;
  [E_{\pm}^{(q)}, F_{\mp}^{(q)}]= \mp U_{\pm}^{(q)}
\nonumber \\
& &  [E_{3}^{(q)}, E_{\pm}^{(q)}]= \pm E_{\pm}^{(q)}
\;\;\;\;\;\;\;\;\;\;
  [F_{3}^{(q)}, F_{\pm}^{(q)}]= \pm F_{\pm}^{(q)}
\nonumber \\
& &  [E_{3}^{(q)}, U_{\pm}^{(q)}]= \pm U_{\pm}^{(q)}
\;\;\;\;\;\;\;\;\;\;
  [F_{3}^{(q)}, U_{\pm}^{(q)}]= \mp U_{\pm}^{(q)}
\nonumber \\
& &  [E_{3}^{(q)}, V_{\pm}^{(q)}]= \pm V_{\pm}^{(q)}
\;\;\;\;\;\;\;\;\;\;
  [F_{3}^{(q)}, V_{\pm}^{(q)}]= \pm V_{\pm}^{(q)}
\nonumber \\
& &  [E_{+}^{(q)}, E_{-}^{(q)}]=-E_{3}^{(q)}
\;\;\;\;\;\;\;\;\;\;
  [F_{+}^{(q)}, F_{-}^{(q)}]= F_{3}^{(q)}
\nonumber \\
& &  [U_{+}^{(q)}, U_{-}^{(q)}]=-(E_{3}^{(q)} - F_{3}^{(q)})
\nonumber \\
& &  [V_{+}^{(q)}, V_{-}^{(q)}]=-(E_{3}^{(q)} + F_{3}^{(q)}).
\end{eqnarray}
and otherwise vanishes.

 Moreover, introducing order-parameter operators
\begin{eqnarray}
\label{N3}
& & \Delta_{+}^{(k)} =\frac{1}{\sqrt 2}(a^+_{k}b^+_{-k}
 -a^+_{-k}b^+_{k}) \;,\;\;\;\;\;
 N_{+}^{(k)} =\frac{1}{\sqrt 2}(a_{k}b^+_{k}
 -a_{-k}b^+_{-k})
\nonumber \\
& & \Delta_{-}^{(k)} =\frac{1}{\sqrt 2}(a_{k}b_{-k}
 -a_{-k}b_{-k}) \;,\;\;\;\;\;
  N_{-}^{(k)} =\frac{1}{\sqrt 2} (a^+_{k}b_{k}
 -a^+_{-k}b_{-k})
\nonumber \\
& & N_{3}^{(k)} =\frac{1}{2} (n^b_{k}-n^b_{-k}
 -n^a_{k}+n^a_{-k})
\end{eqnarray}
and conserved quantity
\begin{equation}
\label{Q}
  Q^{(k)} =\frac{1}{2}(n^a_{k}+n^b_{k}
 -n^a_{-k}-n^b_{-k}-2)
\end{equation}
that obey the commutation relations

\begin{eqnarray}
\label{comutation1}
 & & [N_{\pm}^{(k)}, F_{\mp}^{(k)}]= \pm N_{3}^{(k)}
 \;\;\;\;\;\;\;\;\;\;
  [F_{3}^{(k)}, N_{\pm}^{(k)}]= \pm N_{\pm}^{(k)}
  \nonumber \\
 & &  [N_{\pm}^{(k)}, U_{\pm}^{(k)}]= \mp \Delta_{\pm}^{(k)}
 \;\;\;\;\;\;\;\;\;\;
  [N_{\pm}^{(k)}, V_{\mp}^{(k)}]= \mp \Delta_{\mp}^{(k)}
  \nonumber \\
  & &  [\Delta_{\pm}^{(k)}, E_{\mp}^{(k)}]= \pm N_{3}^{(k)}
  \;\;\;\;\;\;\;\;\;\;
  [\Delta_{\pm}^{(k)}, U_{\mp}^{(k)}]= \pm N_{\pm}^{(k)}
  \nonumber \\
  & &  [\Delta_{\pm}^{(k)}, V_{\mp}^{(k)}]= \mp N_{\mp}^{(k)}
  \;\;\;\;\;\;\;\;\;\;
  [E_{3}^{(k)}, \Delta_{\pm}^{(k)}]= \pm \Delta_{\pm}^{(k)}
  \nonumber \\
  & &  [N_{3}^{(k)}, E_{\pm}^{(k)}]= \mp \Delta_{\pm}^{(k)}
  \;\;\;\;\;\;\;\;\;\;
  [N_{3}^{(k)}, F_{\pm}^{(k)}]= \pm N_{\pm}^{(k)}
  \nonumber \\
  & &  [N_{3}^{(k)}, N_{\pm}^{(k)}]= \pm F_{\pm}^{(k)}
  \;\;\;\;\;\;\;\;\;\;
  [N_{3}^{(k)}, \Delta_{\pm}^{(k)}]= \pm E_{\pm}^{(k)}
  \nonumber \\
  & &  [N_{+}^{(k)}, N_{-}^{(k)}]=F_{3}^{(k)}
  \;\;\;\;\;\;\;\;\;\;\;\;\;
  [N_{\pm}^{(k)}, \Delta_{\pm}^{(k)}]=\pm V_{\pm}^{(k)}
  \nonumber \\
& & [\Delta_{+}^{(k)}, \Delta_{-}^{(k)}]=E_{3}^{(k)}
  \;\;\;\;\;\;\;\;\;\;\;\;\;
 [N_{\pm}^{(k)}, \Delta_{\mp}^{(k)}]=\pm U_{\mp}^{(k)}
\end{eqnarray}

Therefore, we can rewrite $H$ in terms of the generators
of the algebra $SO(3,2)$ and
its order parameter operators as follows:
\begin{eqnarray*}
 H &=& \epsilon_{0}(2E_{3}^{(0)} -1) + g_{\mu}B F^{(0)}_{3}
 +\sum_{k\not=0}(\epsilon_{k}E^{(k)}_{3}+\frac{1}{2}g_{\mu}BF^{(k)}_{3}) \\
 & & +2g_{2}V_0 (U_{+}^{(0)}U_-^{(0)} + V_+^{(0)}V_-^{(0)}
   + E_+^{(0)}E_-^{(0)}) \\
& & +\frac{1}{2}g_{2}V_{0} \sum_{k\not=0}\frac{1}{2^k}
  \{6E_{3}^{(0)}(E_{3}^{(k)} +Q^{(k)})
+2F_3^{(0)}(F_3^{(k)}+ N_{3}^{(k)}) \\
& & + 2F_-^{(0)}(F_+^{(k)}+ N_+^{(k)})
+ 2F_+^{(0)}(F_-^{(k)}+N_-^{(k)})
-3E_{3}^{(k)} \\
& & - 3Q^{(k)}\}
+g_{2}V_{0}\sum_{k\not=0}\frac{1}{\sqrt{k!(-k)!}}
 (U_+^{(0)}U_-^{(k)}
 + U_-^{(0)}U_+^{(k)} \\
& & + V_+^{(0)}V_-^{(k)}
+V_-^{(0)}V_+^{(k)} + E_-^{(0)}E_+^{(k)}
+ E_+^{(0)}E_-^{(k)})
\end{eqnarray*}
Noting that the the algebraic mean-field procedure is a good approximation
to describe condensate and using\cite{solo}

 $$ AB \simeq A\langle B\rangle+\langle A \rangle B-
 \langle A \rangle \langle B\rangle
 $$
The Hamiltonian becomes
\begin{equation}
\label{e8}
 H_{mf}=H^{(0)} + \sum_{k\not=0}H^{(k)} - E_{\ast}
 \;\;\;\;\;\;\;\;\;\;\;\;\;\;\;\;\;\;\;\;\;\;\;\;\;\;\;
\end{equation}
where
\begin{eqnarray}
\label{e9}
H^{(0)} &=& \alpha_0 E^{(0)}_3 + \beta_0 F^{(0)}_3
        +\gamma_0 F^{(0)}_- + \gamma^{\ast}_0 F^{(0)}_+
        +\rho_0 U^{(0)}_- \nonumber \\
  & &   + \rho^{\ast}_0 U^{(0)}_+
        +\sigma_0 V^{(0)}_- + \sigma^{\ast}_0 V^{(0)}_+
        +\tau_0 E^{(0)}_- + \tau^{\ast}_0 E^{(0)}_{+}
\end{eqnarray}
with
\begin{eqnarray*}
\alpha_0 &=& 2\epsilon_{0} + 3g_{2}V_{0}\sum_{k\not=0}
\frac{1}{2^k}(\langle E^{(k)}_{3} \rangle
+ \langle Q^{(k)} \rangle) \\
\beta_0 &=& g_{\mu}B+g_{2}V_{0} \sum_{k\not=0}\frac{1}{2^k}
(\langle F^{(k)}_{3} \rangle
+\langle N^{(k)}_{3} \rangle) \\
\rho_0 &=& g_{2}V_{0}(2\langle U^{(0)}_{+} \rangle
        + \sum_{k\not=0}\frac{1}{\sqrt{k!(-k)!}}
        \langle U^{(k)}_{+} \rangle) \\
\gamma_0 &=& g_{2}V_{0} \sum_{k\not=0} \frac{1}{2^k}
(\langle F^{(k)}_{+} \rangle
+\langle N^{(k)}_{+} \rangle) \\
\sigma_0 &=& g_{2}V_{0} (2\langle V^{(0)}_{+} \rangle
          + \sum_{k\not=0}\frac{1}{\sqrt{k!(-k)!}}
          \langle V^{(k)}_{+} \rangle) \\
\tau_0 &=& g_{2}V_{0} (2\langle E^{(0)}_{+} \rangle
        + \sum_{k\not=0}\frac{1}{\sqrt{k!(-k)!}}
         \langle E^{(k)}_{+} \rangle)
\end{eqnarray*}

\begin{eqnarray}
\label{e10}
H^{(k)} &=& \alpha_k E^{(k)}_3 +(\alpha_{k}-\epsilon_{k})Q^{(k)}
        + (\beta_k-\frac{1}{2}g_{\mu}B) N^{(k)}_{3} \nonumber \\
& &     +\beta_{k} F^{(k)}_3 +\gamma_k (F^{(k)}_- + N^{(k)}_- )
        + \gamma^{\ast}_k (F^{(k)}_+ + N^{(k)}_+) \nonumber \\
& & +\rho_k U^{(k)}_- + \rho^{\ast}_k U^{(k)}_+
    +\sigma_k V^{(k)}_- + \sigma^{\ast}_k V^{(k)}_+  \nonumber \\
& &  +\tau_k E^{(k)}_- + \tau^{\ast}_k E^{(k)}_+
\end{eqnarray}
with

\[ \begin{array}{ll}
\alpha_k = \epsilon_{k} + \frac{g_{2}V_{0}}{2^k}
         (3\langle E^{(0)}_{3} \rangle - \frac{3}{2}) , &
\beta_k = \frac{1}{2} g_{\mu}B+\frac{g_{2}V_{0}}{2^k} \langle F^{(0)}_{3} \rangle , \\

\gamma_k = \frac{g_{2}V_{0}}{2^k} \langle F^{(0)}_{+} \rangle ,&
\rho_k = \frac{g_{2}V_{0}}{\sqrt{k!(-k)!}} \langle U^{(0)}_{+} \rangle ,\\

\sigma_k = \frac{g_{2}V_{0}}{\sqrt{k!(-k)!}} \langle V^{(0)}_{+} \rangle ,&
\tau_k = \frac{g_{2}V_{0}}{\sqrt{k!(-k)!}} \langle E^{(0)}_{+} \rangle .
\end{array}\]

\begin{eqnarray}
\label{E}
E_{\ast}&=&\epsilon_0 +\alpha_{0}\langle E^{(0)}_{3} \rangle
+\beta_{0}\langle F^{(0)}_{3} \rangle
+\gamma_{0}\langle F^{(0)}_{-} \rangle
+\gamma^{ast}_{0}\langle F^{(0)}_{+} \rangle
\nonumber \\
& & +\rho_{0}\langle U^{(0)}_{-} \rangle
         +\sigma_{0}\langle V^{(0)}_{-} \rangle
         +\tau_{0}\langle E^{(0)}_{-} \rangle
         +\sum_{k\not=0}\{\rho_{k}\langle U^{(k)}_{-} \rangle
\nonumber \\
& &        +\sigma_{k}\langle V^{(k)}_{-} \rangle
         +\tau_{k}\langle E^{(k)}_{-} \rangle\}
\end{eqnarray}
Note that the $H_{q}$ $(q=0,k,-k)$ is written in terms of
 $SO(3,2)$ generators and its order
 parameter operators for a given $k$.
 It is known that within the $SO(3,2)$ mean-field picture the energy
 eigenstates are expressed as a direct product of $SO(3,2)$ coherent
 states $\otimes_{q} |\xi_{q} \rangle$. Therefore the eigenstates
 $|\xi \rangle$ can be written as
 \begin{eqnarray}
 \label{e12}
     |\xi \rangle=\otimes_{q}|\xi_{q} \rangle
                 =\otimes_{q}W(\xi_{q})|00 \rangle
\;\;\;\;(q=0,+k,-k)
\end{eqnarray}
where
\begin{eqnarray*}
 W(\xi_{q})&=&\exp\{\xi_{q}(\sqrt{2}\cos{\Theta_{q}}E^{(q)}_{+}
  -\sin{\Theta_{q}}e^{i\Phi_{q}}U^{(q)}_{+} \\
& & +\sin{\Theta_{q}}e^{-i\Phi_{q}}V^{(q)}_{+})
      -H.C.\}
\end{eqnarray*}
with the coherent parameter $ \xi_{q}={r_{k}}e^{i\Psi_{q}}$ and
$|00 \rangle$ is the vacuum state. Using the relations given in
Appendix A, we immediately have
\begin{eqnarray}
\label{e13}
 W^{\dag}(\xi_{q})H_{ q}W(\xi_{ q})
 &=&f_{1}(q) E^{(q)}_{3}
 + f_{2}(q) F^{(q)}_{3}
 + f_{3}(q) E^{(q)}_{+}
 +f^{\ast}_{3}(q) E^{(q)}_{-} \nonumber \\
& & + f_{4}(q) F^{(q)}_{+}
 +f^{\ast}_{4}(q) F^{(q)}_{-}
 +f_{5}(q) U^{(q)}_{+}
  +f^{\ast}_{5}(q) U^{(q)}_{-} \nonumber \\
& &  +f_{6}(q) V^{(q)}_{+}
  +f^{\ast}_{6}(q) V^{(q)}_{-}
  +f_{7}(q) \Delta^{(q)}_{+}
  +f^{\ast}_{7}(q) \Delta^{(q)}_{-} \nonumber \\
& & +f_{8}(q)N^{(q)}_{+}
+f^{\ast}_{8}(q)N^{(q)}_{-}
+f_{9}(q)N^{(q)}_{3}
+f_{10}(q)Q^{(q)}
\end{eqnarray}
where $f_{1}(q), \cdot\cdot\cdot, f_{10}(q)$ are given Appendix B.
Denoting $\sigma_q=|\sigma_q|e^{-i(\Psi_{q}-\Phi_{q})}$,
$\rho_q=|\rho_q|e^{-i(\Psi_{q}+\Phi_{q})}$,
$\tau_q=|\tau_q|e^{-i\Psi_{q}}$, and
$\gamma_q=|\gamma_q|e^{i\Phi_{q}}$, and setting
$f_{3}(q)=f_{4}(q)=f_{5}(q)=f_{6}(q)
=f_{7}(q)=f_{8}(q)=0$, we diagonalized the Hamiltonian $H_q$ as follows:

(1). When $\bf {B}=0$ by direct calculation we find the conditions to have
\begin{eqnarray*}
& &\beta_{q}=|\gamma_{q}|=0,\;\;\;
|\sigma_{q}|=-|\rho_{q}|=\frac{1}{\sqrt{2}}|\tau_{q}|\tan{\Theta_{q}} \\
& &\tanh{2{r_{q}}}=-\frac{\sqrt{2}|\tau_q|}{\alpha_{q}\cos{\Theta_{q}}},
\;\;\;\;\;
E_{q}=\sqrt{{\alpha_{q}}^{2}-2|\tau_{q}|^2\sec^{2}{\Theta_{q}}}
\end{eqnarray*}
and
$$
W^{\dag}(\xi_{q})H_{q}W(\xi_{q})=E_{q}E^{(q)}_{3}+f_{10}(q)Q^{(q)}.
$$
It indicates that in the condensate state of the system there is not
Zeeman effect, however the excitation states produce Zeeman effect
at $k$ and $-k$.

(2). When $\bf {B}\not=0$ we obtain
\begin{eqnarray*}
& &\sin{\Theta_{q}}=0,\;\;\;|\rho_{q}|=|\sigma_{q}|=|\gamma_{q}|=0, \\
& &\beta_{q}=g_{\mu}B \;\;(q=0);
\;\;\; \frac{1}{2} g_{\mu}B \;\;(q\not=0) ,\\
& &\tanh{2{r_{q}}}=-\frac{\sqrt{2}|\tau_q|}{\alpha_{q}\cos{\Theta_{q}}},
\;\;\;
E_{q}=\sqrt{{\alpha_{q}}^{2}-2|\tau_{q}|^2}
\end{eqnarray*}
and
$$
W^{\dag}(\xi_{q})H_{q}W(\xi_{q})=E_{q}E^{(q)}_{3}
+\beta_{q}F^{(q)}_{3}+f_{10}(q)Q^{(q)}.
$$

\section{Super-Poissonian distribution and
correlation functions for the two-component BEC}

The sub-Poissonian photon statistics of light is one of the best
known nonclassical effects. With the rapid development of atom optics,
especially nonclassical motional states of atoms have been generated
in experiments, it is of somewhat importance to investigate nonclassical
effects of atoms. We here discuss the sub-Poissonian distribution of
two-component BEC. Following Mandel\cite{man} the Q parameters
for two-component BEC is introduced:
$$
Q_{a}(0)=\frac {\langle (\Delta {n^{a}_{0}})^2  \rangle}
                     {{\langle n^{a}_{0} \rangle}}-1  \;\;\;\;
Q_{b}(0)=\frac {\langle (\Delta {n^{b}_{0}})^2 \rangle}
                     {{\langle n^{b}_{0} \rangle}}-1
$$
The sub-Poissonian atom statistics exists whenever
$-1 \leq Q_{a(b)}(0)<0$. When $Q_{a(b)}(0)> 0$, the state is
called super-Poissonian while the state with $Q_{a(b)}(0)=0$
is called Poissonian.

Correlations between the two-component BEC hyperfine spin states
of the same atom may be characterized by the second-order correlation
functions:
\begin{eqnarray*}
g^{(2)}_{a}(0)&=&\frac {\langle {a^{+}_{0}}^2 {a_{0}}^2 \rangle}
                     {{\langle a^{+}_{0} a_{0} \rangle}^2}
               =1+\frac{Q_{a}(0)}{\langle{n_{0}^{a}}\rangle} \\
g^{(2)}_{b}(0)&=&\frac {\langle {b^{+}_{0}}^2 {b_{0}}^2 \rangle}
                     {{\langle b^{+}_{0} b_{0} \rangle}^2}
               =1+\frac{Q_{b}(0)}{\langle{n_{0}^{b}}\rangle} \\
g^{(2)}_{ab}(0)&=&\frac {\langle {n^{a}_{0}}{n^{b}_{0}} \rangle}
                     {{\langle n^{a}_{0} \rangle}{\langle n^{b}_{0}} \rangle}
\end{eqnarray*}
where $g^{(2)}_{ab}(0)=1$ for uncorrelated
states; $ g^{(2)}_{ab}(0) > 1 $ for correlated states
and $ g^{(2)}_{ab}(0) < 1 $ for anticorrelated
states.
For a system consisting of two-component BEC, there is the Cauchy-Schwartz
inequality (CSI)\cite{agar}
\begin{equation}
\label{csi}
[g^{(2)}_{ab}(0)]^{2}\leq g^{(2)}_{a}(0)g^{(2)}_{b}(0)
\end{equation}
Reid and Walls\cite{raid} showed that violations of the CSI can be accompanied
by the violations of Bell's inequality. If the inequality (\ref{csi}) is
violated, the correlations between two components are called nonclassical
correlations which can be characterized by
$$
I(0)=\frac{\sqrt{g^{(2)}_{a}(0)g^{(2)}_{b}(0)}}{g^{(2)}_{ab}(0)}-1
$$
which is negative if the inequality (\ref{csi})is violated.
For the states $\mid{\xi}\rangle$ given in eq(\ref{e12}), the correlation
functions
$g^{(2)}_{a}(0) = g^{(2)}_{b}(0)=2$ and
$g^{(2)}_{ab}(0)=1+\coth^{2}{r_0}\cos^{2}{\Theta_0}$,
that do not agree with the experimental results,
which seem to indicate that $g^{(2)}(0)$ are not
exactly equal to 1, slightly larger than 1. It is
easy to show that $g^{(2)}_{a}=g^{(2)}_{b}=g^{(2)}_{ab}=1$
in the state $D(\alpha,\beta)\mid{0}\rangle$ (D state) if the mean
density $\langle{n^{a}_0}\rangle$  and $\langle{n^{b}_{0}}\rangle$
are a large numbers, where $D(z^{a}_{q},z^{b}_{q})
=\exp(z^{a}_{q}a^+_q+z^{b}_{q}b^{+}_{q}-h.c.)$.

These considerations motivate our attempt to generalize $|\xi \rangle$ to
$|\xi,z^{a},z^{b} \rangle$,
\begin{equation}
\label{zazb1}
|\xi,z^{a},z^{b}\rangle=|\xi_{0},z^{a}_{0},z^{b}_{0}\rangle
      \otimes_{k\not=0}|\xi_{k},z^{a}_{k},z^{b}_{k}\rangle,
\end{equation}
by introducing the further definitions
\begin{equation}
\label{z0z0}
|\xi_{0},z^{a}_{0},z^{b}_{0}\rangle=D(z^{a}_{0},z^{b}_{0})|\xi_{0}\rangle,\;\;\;
|\xi_{k},z^{a}_{k},z^{b}_{k}\rangle=D(z^{a}_{k},z^{b}_{k})|\xi_{k}\rangle,
\end{equation}
where $D(z^{a}_{q},z^{b}_{q})=\exp(z^{a}_{q}a^+_q+z^{b}_{q}b^{+}_{q}-h.c.)$,
$q=0,k,-k$.
We now describe the BEC states by $|\xi,z^{a},z^{b}\rangle$ where
\begin{equation}
\label{zazb2}
|\xi,z^{a},z^{b}\rangle=\otimes_{q}|\xi_{q},z^{a}_{q},z^{b}_{q}\rangle
=\otimes_{q}D(z^{a}_{q},z^{b}_{q})W(\xi_{q})|00 \rangle,\\
(q=0,\pm1,\pm2,...).
\end{equation}
For convenience, we refer to the state $|\xi,z^{a},z^{b}\rangle$ as a DW
state, the DW operator being similar to, but not identical with, what
produces a squeezed state in quantum optics.

\begin{eqnarray}
\label{ww}
 W^{+}(\xi_q) a_q W(\xi_q)&=& a_q\cosh{{r_{q}}}+(\cos{\Theta_{q}}b^{+}_{-q}
                          -\sin{\Theta_{q}}e^{i\Phi_{q}}a^{+}_{-q})
                          e^{i\Psi_{q}}\sinh{{r_{q}}}  \nonumber \\
 W^{+}(\xi_q) b_q W(\xi_q)&=& b_q\cosh{{r_{q}}}+(\cos{\Theta_{q}}a^{+}_{-q}
                          +\sin{\Theta_{q}}e^{-i\Phi_{q}}b^{+}_{-q})
                          e^{i\Psi_{q}}\sinh{{r_{q}}}
\end{eqnarray}

We obtain the following mean values in the DW state:
\begin{eqnarray*}
\langle n^{a}_{0} \rangle
&=&|{z^{a}_{0}}|^2 + \sinh^2{r_0},\;\;\;\;\;
\langle n^{b}_{0} \rangle=|{z^{b}_{0}}|^2
+ \sinh^2{r_0} \\
\langle {({n^{a}_{0}})^2} \rangle
&=& (|{z^{a}_{0}}|^2
+ \sinh^{2}{r_0})^2 + \cosh^{2}{r_0}(|{z^{a}_{0}}|^2
+ \sinh^{2}{r_0}) + |z^{a}_{0}|^2 \sinh^{2}{r_0}\\
& & -\frac{1}{2}\sinh{2r_0}\sin{\Theta_{0}}
[({{z^{a}_{0}}^{\ast}})^2e^{i(\Phi_{0}+\Psi_{0})}
+({z^{a}_{0}})^2e^{-i(\Phi_{0}+\Psi_{0})}] \\
\langle {({n^{b}_{0}})^2} \rangle
&=& (|{z^{b}_{0}}|^2
+ \sinh^2{r_0})^2 + \cosh^2{r_0}(|{z^{b}_{0}}|^2
+ \sinh^2{r_0}) + |z^{b}_{0}|^2 \sinh^2{r_0} \\
& & +\frac{1}{2}\sinh{2r_0}\sin{\Theta_{0}}
[({{z^{b}_{0}}^{\ast}})^2e^{i(\Psi_{0}-\Phi_{0})}
+({z^{b}_{0}})^2e^{i(\Phi_{0}-\Psi_{0})}] \\
\langle {n^{a}_{0}n^{b}_{0}}\rangle
&=& (|{z^{a}_{0}}|^2+\sinh^2{r_0})
(|{z^{b}_{0}}|^2 + \sinh^2{r_0})
+\cosh^{2}{r_0} \sinh^{2}{r_0} \cos^{2}{\Theta_{0}} \\
& &+\frac{1}{2}\sinh{2r_0}\cos{\Theta_{0}}
(e^{i\Psi_{0}}{z^a_0}^{\ast}{z^b_0}^{\ast}
+e^{-i\Psi_{0}}{z^a_0}{z^b_0})\\
& &-2{z^a_0}{z^b_0}^{\ast} \sinh^2{r_0} \sin{2\Theta_{0}}
\sin{\Phi_{0}}
\end{eqnarray*}
If we take $z^{a}_{0}=|z^{a}_{0}|\exp(i\delta_0)$ and
$z^{b}_{0}=|z^{b}_{0}|\exp(i\delta_0)$, then the value $g^{(2)}_{a}(0)$,
$g^{(2)}_{b}(0)$ and $g^{(2)}_{ab}(0)$ for the DW state are
\begin{eqnarray}
\label{ggg1}
g^{(2)}_{a}(0)&=&\frac{|z^{a}_0|^2 \sinh^{2}{r_0}[1-2\coth{r_0}\sin{\Theta_{0}}
\cos(\Psi_{0}+\Phi_{0}-2\delta_0)]}{(|{z^{a}_{0}}|^2 + \sinh^2{r_0})^2}
\nonumber \\
& &+1+\frac{\sinh^{2}{r_0}}{|{z^{a}_{0}}|^2+\sinh^{2}{r_0}} \\
\label{ggg2}
g^{(2)}_{b}(0)&=&\frac{|z^{b}_0|^2 \sinh^{2}{r_0}[1+2\coth{r_0}\sin{\Theta_{0}}
\cos(\Psi_{0}-\Phi_{0}-2\delta_0)]}{(|{z^{b}_{0}}|^2 + \sinh^2{r_0})^2}
\nonumber \\
& &+1+\frac{\sinh^{2}{r_0}}{|{z^{b}_{0}}|^2+\sinh^{2}{r_0}} \\
\label{ggg3}
g^{(2)}_{ab}(0)&=&1+\frac{|z^{a}_0||z^{b}_0|\sinh{2r_0}
[\cos{(\Psi_{0}-2\delta_0)}
-\frac{1}{2}\tanh{r_0}\sin{2\Theta_{0}}
\sin{\Phi_{0}}]}{(|{z^{a}_{0}}|^2 + \sinh^2{r_0})
(|{z^{b}_{0}}|^2 + \sinh^2{r_0})} \nonumber \\
& & + \frac{\sinh^2{2{r_0}}\cos^2{\Theta_{0}}}
{4(|{z^{a}_{0}}|^2 + \sinh^2{r_0})
(|{z^{b}_{0}}|^2 + \sinh^2{r_0})}
\end{eqnarray}
Parelleling to the above section we also distinguish two cases:

(1). If $\bf {B}=0$ then $\sin{\Theta_{0}}$ may take arbitrary value,
therefore the above relations eq.(\ref{ggg1})-(\ref{ggg3}) are unchanged.
According to the above relations eq.(\ref{ggg1})-(\ref{ggg3}) we conclude
that the distribution of the the two-component BEC is uncertain.

(2). If there exists the magnetic field $\bf B$ then
$\sin{\Theta_{0}}=0$. When choosing
$|z^{a}_{0}|=|z^{b}_{0}|=z_{0}$
we obtain the following relations:
\begin{eqnarray}
\label{ga}
 g^{(2)}_{a}(0)&=&g^{(2)}_{b}(0)
=1+\frac{\sinh^{2}{r_0}}{{z_{0}}^2+\sinh^{2}{r_0}}
+\frac{z_0^2 \sinh^{2}{r_0}}{({z_{0}}^2 + \sinh^2{r_0})^2} \\
\label{Qa}
 Q_{a}(0)&=&Q_{b}(0)
=\sinh^{2}{r_0}
+\frac{z_0^2 \sinh^{2}{r_0}}
{{z_{0}}^2 + \sinh^2{r_0}} \\
\label{gab}
 g^{(2)}_{ab}(0)&=&1+\frac{z_0^{2}\sinh{2r_0}
\cos{(\Psi_{0}-2\delta_0)} + \frac{1}{4}\sinh^2{2{r_0}}}
{({z_{0}}^2 + \sinh^2{r_0})^{2}} \\
\label{I00}
 I(0)&=&\frac{\sinh^{2}{r_0}(2z^{2}_{0}-1)-z^{2}_{0}\sinh{2r_0}
\cos{(\Psi_{0}-2\delta_{0})}}{(z^{2}_{0}+\sinh^{2}{r_0})^{2}
+\frac{1}{4}\sinh^{2}{2r_0}+z^{2}_{0}\sinh{2r_0}
\cos{(\Psi_{0}-2\delta_{0})}}
\end{eqnarray}
From eq.({\ref{ga}}) and eq.(\ref{Qa}) it immediately follows that
$1<g^{(2)}_{{a}(b)}(0)<2$ and $Q_{{a}(b)}(0)>0$. It indicates that
the two-component BEC obey super-Poissonian distribution. However,
the properties of $g^{(2)}_{ab}(0)$ and $I(0)$ depend on the value
of $\cos{(\Psi_{0}-2\delta_{0})}$.

\section{Conclusion}
 In this paper we have diagonalized the system of a
 two-component BEC base on the $SO(3,2)$ spectrum-generating
algebra structure for the mean field Hamiltonian and
shown that the eigenstate is related to $SO(3,2)$-coherent
state. Also we find that a two-component BEC associated with
DW state satisfy uniquely super-Poissonian distribution
in the fixed magnetic field along the $z$-direction, but
as the magnetic field disappears the distribution wil become
uncertain. Therefore the DW state will provide better fits
to the experimental results on the correlation function
associated with the BEC state.

\vspace{3mm}

{ACKNOWLEDGMENT}

This work was partially supported by the National Natural Science Foundation
of China.

\vspace{3mm}

{APPENDIX  A}
\begin{eqnarray*}
   W^{\dag}(\xi_{q})E^{(q)}_{\pm}W(\xi_{q})
   &=& E^{(q)}_{\pm}\cosh^{2}{r_{q}} -\{\frac{1}{\sqrt{2}}
   \sin{2\Theta_{q}}(e^{\mp i(2\Psi_{q}+\Phi_{q})}U^{(q)}_{\mp} \\
& & -e^{\mp i(2\Psi_{q}-\Phi_{q})}V^{(q)}_{\mp})
-e^{\mp i2\Psi_{q}}\cos{2\Theta_{q}}E^{(q)}_\mp\}
\times \sinh^2{r_{q}}\\
& & +\{\frac{1}{\sqrt{2}}e^{\mp i\Psi_{q}}\cos{\Theta_{q}}E^{(q)}_{3}
+ \frac{1}{2} \sin{\Theta_{q}}(e^{\mp i(\Psi_{q}-\Phi_{q})}F^{(q)}_{\mp} \\
& & -e^{\mp i(\Psi_{q}+\Phi_{q})}F^{(q)}_{\pm})\} \times \sinh2{r_{q}},
\end{eqnarray*}
\begin{eqnarray*}
   W^{\dag}(\xi_{q})U^{(q)}_{\pm}W(\xi_{q})
   &=& U^{(q)}_{\pm}\cosh^{2}{r_{q}} -\{\frac{1}{\sqrt{2}}
   \sin{2\Theta_{q}}e^{\mp i(2\Psi_{q}+\Phi_{q})}E^{(q)}_{\mp} \\
& & -e^{\mp i2(\Psi_{q}+\Phi_{q})}\sin^2{\Theta_{q}}U^{(q)}_{\mp}
-e^{\mp i2\Psi_{q}}\cos^2{\Theta_{q}}V^{(q)}_{\mp}\} \\
& & \times \sinh^2{r_{q}}
-\{\frac{1}{2}e^{\mp i(\Psi_{q}+\Phi_{q})}\sin{\Theta_{q}}[E^{(q)}_{3}
-F^{(q)}_{3}] \\
& & - \frac{1}{\sqrt{2}} e^{\mp i\Psi_{q}}\cos{\Theta_{q}}F^{(q)}_{\mp})]
\times \sinh2{r_{q}}
\end{eqnarray*}
\begin{eqnarray*}
   W^{\dag}(\xi_{q})V^{(q)}_{\pm}W(\xi_{q})
  &=& V^{(q)}_{\pm}\cosh^{2}{r_{q}} + \{\frac{1}{\sqrt{2}}
   \sin{2\Theta_{q}}e^{\mp i(2\Psi_{q}-\Phi_{q})}E^{(q)}_{\mp} \\
& & +e^{\mp i2(\Psi_{q}-\Phi_{q})}\sin^2{\Theta_{q}}V^{(q)}_{\mp}
+e^{\mp i2\Psi_{q}}\cos^2{\Theta_{q}}U^{(q)}_{\mp}\}  \\
& & \times \sinh^2{r_{q}}
+\{\frac{1}{2}e^{\mp i(\Psi_{q}-\Phi_{q})}\sin{\Theta_{q}}[E^{(q)}_{3}
+F^{(q)}_{3}] \\
& & + \frac{1}{\sqrt{2}} e^{\mp i\Psi_{q}}\cos{\Theta_{q}}F^{(q)}_{\pm})\}
\times \sinh2{r_{q}},
\end{eqnarray*}
\begin{eqnarray*}
     W^{\dag}(\xi_{q})F^{(q)}_{\pm}W(\xi_{q})
   &=& \{\frac{1}{\sqrt{2}}\sin{2\Theta_{q}}
    e^{\pm{i\Phi_{q}}}F^{(q)}_{3}
     -\sin^{2}{\Theta_{q}}e^{\pm {2i\Phi_{q}}}F^{(q)}_{\mp}\}
    \times \sinh^{2}{{r_{q}}} \\
  & &  + \{\frac{1}{2}\sin{\Theta_{q}}[e^{\mp{i(\Psi_{q}-\Phi_{q})}}E^{(q)}_{\mp}
    -e^{\pm{i(\Psi_{q}+\Phi_{q})}}E^{(q)}_{\pm}] \\
& &   +\frac{1}{\sqrt{2}}\cos{\Theta_{q}}[e^{\mp{i\Psi_{q}}}U^{(q)}_{\mp}
   +e^{\pm{i\Psi_{q}}}V^{(q)}_{\pm}]\}\times \sinh2r_{q} \\
 & & +(\cosh^{2}{{r_{q}}}+\cos^{2}{\Theta_{q}}\sinh^{2}{{r_{q}}})F^{(q)}_{\pm}
\end{eqnarray*}
\begin{eqnarray*}
   W^{\dag}(\xi_{q})E^{(q)}_{3}W(\xi_{q})
  &=& E^{(q)}_{3}\cosh{2}{r_{q}}
   + \frac{1}{\sqrt{2}} \{e^{i\Psi_{q}}E^{(q)}_{+}
   + e^{-i\Psi_{q}}E^{(q)}_{-}\} \\
& &  \times \sinh2{r_{q}} \cos{\Theta_{q}}
 -\frac{1}{2}\{e^{i(\Psi_{q}+\Phi_{q})}U^{(q)}_{+}
+e^{-i(\Psi_{q}+\Phi_{q})}U^{(q)}_{-}\\
& &-e^{i(\Psi_{q}-\Phi_{q})}V^{(q)}_{+}
-e^{-i(\Psi_{q}-\Phi_{q})}V^{(q)}_{-}\}
\times \sinh2{r_{q}} \sin{\Theta_{q}}
\end{eqnarray*}
\begin{eqnarray*}
  W^{\dag}(\xi_{q})F^{(q)}_{3}W(\xi_{q})
   &=& \frac{1}{\sqrt{2}}[e^{-i\Phi_{q}}F^{(q)}_{+}
   + e^{i\Phi_{q}}F^{(q)}_{-}]
   \times \sinh^{2}{r_{q}} \sin{2\Theta_{q}}\\
& & +\frac{1}{2}\{e^{i(\Psi_{q}+\Phi_{q})}U^{(q)}_{+}
+e^{-i(\Psi_{q}+\Phi_{q})}U^{(q)}_{-}\\
& & +e^{i(\Psi_{q}-\Phi_{q})}V^{(q)}_{+}
+e^{-i(\Psi_{q}-\Phi_{q})}V^{(q)}_{-}\}
\times \sinh2{r_{q}} \sin{\Theta_{q}}\\
& & +(1+2\sin^2{\Theta_{q}}\sinh^{2}{r_{q}})F^{(q)}_{3}
\end{eqnarray*}
\begin{eqnarray*}
W^{\dag}(\xi_{q})N^{(q)}_{\pm}W(\xi_{q})
&=& [e^{\pm i2\Phi_{k}}N^{(q)}_{\mp}
-\frac{1}{\sqrt{2}}\sin{2\Theta_{q}}e^{\pm i\Phi_{q}}N^{(q)}_{3}]
\times \sinh^{2}{r_{q}} \\
& & +\frac{1}{2}[e^{\mp i(\Psi_{q}-\Phi_{q})}\Delta^{(k)}_{\mp}
+e^{\pm i(\Psi_{q}+\Phi_{q})}\Delta^{(q)}_{\pm}]
\times \sinh{2r_q}\\
& & +(1+\sin^{2}{\Theta_{q}}\sinh^{2}{r_{q}})N^{(q)}_{\pm}
\end{eqnarray*}
\begin{eqnarray*}
W^{\dag}(\xi_{q})N^{(q)}_{3}W(\xi_{q})
&=& -\frac{1}{\sqrt{2}}[e^{-i\Phi_{q}}N^{(q)}_{+}
+e^{i\Phi_{q}}N^{(q)}_{-}]
\times \sinh^{2}{r_{q}}\sin{2\Theta_{q}} \\
& &-\frac{1}{\sqrt{2}}[e^{-i\Psi_{q}}\Delta^{(q)}_{-}
+e^{i\Theta_{q}}\Delta^{(q)}_{+})
\times \sinh{2{r_{q}}}\cos{\Theta_{q}} \\
& & +(1+2\cos^{2}{\Theta_{k}}\sinh^{2}{r_{q}})N^{(K)}_{3}
\end{eqnarray*}

\vspace{3mm}

{APPENDIX  B}
\begin{eqnarray*}
  f_{1}(q) &=&\alpha_{q}\cosh{2{r_{q}}}
  + (\sigma_{q}e^{i(\Psi_{q}-\Phi_{q})}
  +\sigma^{\ast}_{q}e^{i(\Phi_{q}-\Psi_{q})} \\
& & -\rho_{q}e^{i(\Psi_{q}+\Phi_{q})}
  -\rho^{\ast}_{q}e^{-i(\Phi_{q}+\Psi_{q})})
  \times \frac{1}{2}\sinh{2{r_{q}}}\sin{\Theta_{q}} \\
& & +\frac{1}{\sqrt{2}}(\tau_{q}e^{i\Psi_{q}}
  +\tau^{\ast}_{q}e^{-i\Psi_{q}})
  \times\sinh{2{r_{k}}}\cos{\Theta_{q}} \\
  f_{2}(q)&=&\beta_{q}(1+2\sin^{2}{\Theta_{q}}\sinh^{2}{{r_{q}}})
  +\frac{1}{2}(\sigma_{q}e^{i(\Psi_{q}-\Phi_{q})}
  +\sigma^{\ast}_{q}e^{i(\Phi_{q}-\Psi_{q})} \\
& &  +\rho_{q}e^{i(\Psi_{q}+\Phi_{q})}
  +\rho^{\ast}_{q}e^{-i(\Phi_{q}+\Psi_{q})})
  \times \sinh{2{r_{q}}}\sin{\Theta_{q}} \\
& &  +\frac{1}{\sqrt{2}}\sinh^{2}{{r_{q}}}\sin{2\Theta_{q}}
  (\gamma_{q}e^{-i\Phi_{q}}
  +\gamma^{\ast}_{q}e^{i\Phi_{q}})  \\
  f_{3}(q) &=& \frac{1}{2}(\gamma_{q}e^{i(\Psi_{q}-\Phi_{q})}
   -\gamma^{\ast}_{q}e^{i(\Psi_{q}+\Phi_{q})})
  \sinh{2{r_{q}}}\sin{\Theta_{q}}\\
& &  +\tau^{\ast}_{q}\cosh^{2}{{r_{q}}}
  + \tau_{q}\sinh^{2}{{r_{q}}}\cos{2\Theta_{q}}e^{i2\Psi_{q}}\\
& & +\frac{1}{\sqrt{2}}(\sigma_{q}e^{i(2\Psi_{q}-\Phi_{q})}
  -\rho_{q}e^{i(2\Psi_{q}+\Phi_{q})})
  \sinh^{2}{{r_{q}}}\sin{2\Theta_{q}}\\
& & +\frac{1}{\sqrt{2}}\alpha_{q}\sinh{2{r_{q}}}
  \cos{\Theta_{q}}e^{i\Psi_{q}} \\
  f_{4}(q) &=& \gamma^{\ast}_{q}(\cosh^{2}{r_{q}}
      +\cos^{2}{\Theta_{q}}\sinh^{2}{r_{q}})
      -\gamma_{q}\sinh^{2}{{r_{q}}}
      \sin^{2}{\Theta_{q}}e^{-i2\Phi_{q}}\\
& & +\frac{1}{\sqrt{2}}(\rho_{q}e^{i\Psi_{q}}
  +\sigma^{\ast}_{q}e^{-i\Psi_{q}})
  \sinh{2{r_{q}}}\cos{\Theta_{q}}\\
& & +\frac{1}{2}(\tau_{q}e^{i(\Psi_{q}-\Phi_{q})}
  -\tau^{\ast}_{q}e^{-i(\Psi_{q}+\Phi_{q})})
 \sinh{2{r_{q}}}\sin{\Theta_{q}}\\
& &+\frac{1}{\sqrt{2}} \beta_{q}\sinh^{2}{{r_{q}}}
  \sin{2\Theta_{q}}e^{-i\Phi_{q}} \\
  f_{5}(q) &=& \frac{1}{2}(\beta_{q}-\alpha_{q})
  \sinh{2{r_{q}}}\sin{\Theta_{q}}e^{i(\Psi_{q}+\Phi_{q})}
   +\rho^{\ast}_{q}\cosh^{2}{{r_{k}}} \\
& & +\frac{1}{\sqrt{2}}\gamma_{q}\sinh{2{r_{q}}}\cos{\Theta_{q}}e^{i\Psi_{q}}
+\{\rho_{q}\sin^{2}{\Theta_{q}}e^{i2(\Psi_{q}+\Phi_{q})}\\
& & + \sigma_{q}\cos^{2}{\Theta_{q}}e^{i2\Psi_{q}}
-\frac{1}{\sqrt{2}}\tau_{q}\sin{2\Theta_{q}}
e^{i(2\Psi_{q}+\Phi_{q})}\}\times \sinh^{2}{r_q} \\
  f_{6}(q) &=& \frac{1}{2}(\alpha_{q}+\beta_{q})
  \sinh{2{r_{q}}}\sin{\Theta_{q}}e^{i(\Psi_{q}-\Phi_{q})}
  +\sigma^{\ast}_{q}\cosh^{2}{{r_{q}}} \\
& & +\frac{1}{\sqrt{2}}\gamma^{\ast}_{q}
\sinh{2{r_{q}}}\cos{\Theta_{q}}e^{i\Psi_{q}}
 +\{\rho_{q}\cos^{2}{\Theta_{q}}e^{i2\Psi_{q}} \\
& & + \sigma_{q}\sin^{2}{\Theta_{q}}e^{i2(\Psi_{q}-\Phi_{q})}
+\frac{1}{\sqrt{2}}\tau_{q}\sin{2\Theta_{q}}
e^{i(2\Psi_{q}-\Phi_{q})}\} \times \sinh^{2}{r_q} \\
f_{7}(q) &=& -\frac{1}{\sqrt{2}}{\beta'}_{q}\sinh{2r_{q}}
\cos{\Theta_{q}}e^{i\Psi_q}
+\frac{1}{2}(\gamma_{q}e^{-i\Phi_{q}}
+\gamma^{\ast}_{q}e^{i\Phi_{q}})
\times \sinh{2r_{q}}e^{i\Psi_{q}} \\
f_{8}(q) &=& -\frac{1}{\sqrt{2}}{\beta'}_{q}\sinh^{2}r_{q}
\sin{2\Theta_{q}}e^{-i\Phi_{q}}
+\gamma_{q}\sinh^{2}r_{q}\sin^{2}{\Theta_{q}}e^{-2i\Phi_{q}} \\
& & +\gamma^{\ast}_{q}(1+\sinh^{2}r_{q}\sin^{2}{\Theta_{q}}) \\
f_{9}(q) &=& {\beta'}_{q}
(1+2\sinh^{2}r_{q}\cos^{2}{\Theta_{q}})
-\frac{1}{\sqrt{2}}(\gamma_{q}e^{-i\Phi_{q}}
+\gamma^{\ast}_{q}e^{i\Phi_{q}})\sinh^{2}r_{q}\sin{2\Theta_{q}} \\
f_{10}(q) &=& 0 (if q=0); \alpha_{q}-\epsilon_{q}(if q\not=0) \\
\beta'_{q} &=& 0 (if q=0); \beta_{q}-\frac{1}{2}g_{\mu}B (if q\not=0).
\end{eqnarray*}

\end{titlepage}
\end{document}